\documentclass[letterpaper,twocolumn,10pt]{article}
\usepackage{usenix2019_v3}

\usepackage{tikz}
\usepackage{amsmath}

\usepackage{filecontents}

\usepackage{kotex}
\usepackage{enumitem}
\usepackage[normalem]{ulem}

\usepackage{booktabs}
\usepackage{multirow}
\usepackage{ulem} 

\usepackage{amssymb}
\usepackage{comment}

\newcommand{\sys}{{StorInfer}}
\begin{document}

\date{}

\title{\LARGE \bf Accelerating LLM Inference with Precomputed Query Storage}

\author{
{\rm Jay H. Park\textsuperscript{†*}, Youngju Cho\textsuperscript{†*}, Choungsol Lee\textsuperscript{†}, Moonwook Oh\textsuperscript{†}, and Euiseong Seo\textsuperscript{‡§}}\\
\rm \textsuperscript{†}Samsung Electronics, \textsuperscript{‡}Sungkyunkwan University\\
\rm \{tino.park, yj93.cho, cs89.lee, mw.oh\}@samsung.com, euiseong@skku.edu
} 

\maketitle

\begingroup
\renewcommand\thefootnote{}\footnotetext{%
\textsuperscript{*} Equal contribution \textsuperscript{§} Corresponding author}
\endgroup

\begin{abstract}

Large language model (LLM) inference often suffers from high latency, particularly in resource-constrained environments such as on-device or edge deployments. To address this challenge, we present {\it \sys{}}, a novel storage-assisted LLM inference system that accelerates response time by precomputing and storing predictable query–response pairs offline. When a user query semantically matches a precomputed query, \sys{} bypasses expensive GPU inference and instantly returns the stored response, significantly reducing latency and compute costs. To maximize coverage and effectiveness, \sys{} employs an LLM-driven generator that adaptively produces diverse and deduplicated queries based on a given knowledge base. This is achieved via two techniques: adaptive query masking, which prevents regeneration of similar queries, and adaptive sampling, which dynamically tunes generation parameters to promote semantic diversity. The resulting query–response pairs are embedded and indexed using a disk-backed vector database to enable fast, similarity-based retrieval at runtime. Using this approach, we generated 150K unique precomputed pairs (taking up to 830 MB of storage space), achieving up to 17.3\% latency reduction with no loss in response quality. Our evaluation across multiple QA datasets demonstrates the practicality and scalability of storage-assisted inference, especially in scenarios with predictable query distributions. \sys{} highlights a promising direction in leveraging storage as a primary enabler for efficient, low-latency LLM deployment.

\end{abstract}

\section{Introduction}
\label{sec:intro}

Large language models (LLMs) have become foundational components across a wide range of applications, owing to their strong capabilities in natural language understanding and generation~\cite{devlin2019bertpretrainingdeepbidirectional, vaswani2017attention, NEURIPS2020_1457c0d6}. They are prominently used in natural language processing tasks such as document summarization, code generation, and speech recognition~\cite{zaheer2020bigbird, chen2021evaluatinglargelanguagemodels, radford2023robust}.

At the same time, there is growing interest in deploying LLMs on on-device systems and edge environments, where computational resources—such as GPU availability, memory bandwidth, and power budget—are significantly constrained~\cite{lin2024awq, articleManufacturing}. These settings require real-time inference capabilities (e.g., for voice assistants or interactive agents), yet lack the capacity to support full-scale LLMs. As a result, inference is often carried out using lightweight models, which frequently leads to degraded output quality or suboptimal response time. These challenges underscore the need for novel strategies that reduce computational demands while preserving high-quality responses.

To address the computational overhead of LLM inference, prior work has primarily adopted cache-based approaches that aim to reuse inference outcomes from previous queries~\cite{openai_promptcaching, deepseek_contextcaching, bang2023gptcache}. However, these approaches are vulnerable to the cold start problem and are only effective when the incoming input exactly matches previously cached content.

In this paper, we propose the {\it \sys{}} system that leverages storage capacity to improve the performance of LLM inference. \sys{} precomputes anticipated queries and their corresponding responses—generated using a high-quality LLM—during an offline stage and stores them in persistent storage. During online inference, \sys{} compares the embedding vector of the incoming query with those of the precomputed queries. If a precomputed query is found whose embedding is within a predefined similarity threshold, the corresponding response is immediately retrieved and returned. If no sufficiently similar precomputed query is found, \sys{} falls back to on-device inference using a lightweight LLM suited for resource-constrained environments.

This study introduces a novel approach to optimizing LLM inference by proactively predicting user queries and precomputing query-response pairs in a dedicated storage system. To realize the proposed approach, we explored the following key research questions (RQs):
\begin{itemize}
  \item \textbf{RQ1.} Can we accelerate LLM inference using a storage system without compromising response quality?
  \item \textbf{RQ2.} Can we generate diverse precomputed queries that effectively anticipate user queries?
  \item \textbf{RQ3.} Does this performance improvement scale proportionally with increased storage usage?
\end{itemize}

To address \textbf{RQ1}, we propose \sys{}, a system that proactively stores precomputed query–response pairs and seamlessly integrates them into LLM inference workflows.
To answer \textbf{RQ2} and \textbf{RQ3}, \sys{} includes a Query Generator module that automatically generates diverse queries and analyzes whether performance improvements scale proportionally with increased storage usage.

Our approach is particularly effective in cases where the distribution of LLM queries is narrow or predictable. For instance, voice recognition systems in vehicle infotainment typically support a limited set of command patterns~\cite{Astuti_Tan_Solihin_Vincent_Michael_2021}. Similarly, retrieval-augmented generation (RAG) systems often operate over a fixed corpus, resulting in semantically repetitive query patterns. In these settings, where input variability is limited, it becomes feasible to anticipate user queries and precompute high-quality responses in advance.
\section{Background and Motivation}
\label{sec:background}
{\bf LLM Inference.}
Large Language Models (LLMs) generate human-like text by predicting the next token based on input sequences.
Inference typically involves two phases: prefill and decode.
During the prefill phase, the entire input prompt is processed, and key-value (KV) cache entries for the attention mechanism are generated — a process that is generally compute-intensive.
The subsequent decode phase generates tokens sequentially, using the KV cache to avoid redundant computation. 
Still, it can be costly for long responses.
Additionally, the decode phase output is influenced by sampling parameters, such as temperature, which controls the randomness of token selection; lower values yield more deterministic responses, while higher values promote greater diversity.

{\bf Vector Search.}
The effective management and retrieval of unstructured data such as text, images, and audio relies on techniques that represent these elements as vectors for similarity-based operations.
Common similarity metrics include maximum inner product search (MIPS)~\cite{shrivastava2014asymmetriclshalshsublinear}, Cosine similarity~\cite{deerwester1990indexing}, and Euclidean distance~\cite{johnson2017billionscalesimilaritysearchgpus}.
For efficient searching within large-scale vector datasets, approximate nearest neighbor (ANN) algorithms are typically employed~\cite{AGATONOVICKUSTRIN2000717}.
In this work, we use MIPS as our primary similarity metric. To enable efficient large-scale search, we adopt DiskANN~\cite{NEURIPS2019_09853c7f}, an ANN-based vector search library.

\subsection{Limitations of Caching in LLM} 
{\bf Prefix Caching.} 
To enhance LLM performance and reduce latency, previously computed KV cache entries for common prompt prefixes are reused.
Building on this idea, OpenAI has introduced Prompt Caching~\cite{openai_promptcaching}, while DeepSeek has implemented Context Caching~\cite{deepseek_contextcaching} as part of their service.
However, the cache may be invalidated with minor changes to prompts, potentially limiting its application depending on the service environment.
However, even minor changes to the prompt can invalidate the cache, and prefix caching is limited to optimizing only the prefill phase.
The latency of the decode phase, where tokens are generated, remains unaffected.
To optimize end-to-end latency, a new approach beyond conventional caching mechanisms is required.

{\bf Query Caching.} 
Another approach for optimizing LLM performance involves caching queries and responses rather than KV caching.
GPTCache~\cite{bang2023gptcache} stores previously answered queries and responses from the LLM, utilizing them to reduce repetitive computations when identical or similar queries are encountered.
Unlike prefix caching, this technique can return cached responses for sufficiently similar inputs, even when prompts are not identical, though it may still suffer from cold start limitations.
Hence, there is a need to go beyond simply caching historical data, by predicting likely future queries and precomputing their responses in advance.
These precomputed query methodologies represent an important research direction that can further improve cache hit rates and enhance response speeds for new queries.



\subsection{Precomputing: Novel Possibilities}
Precomputing represents a promising approach that opens up novel possibilities for improving the efficiency of LLM inference.
By anticipating user queries and precomputing corresponding responses in advance, the system can replace costly inference operations with lightweight retrieval, thereby achieving near-instantaneous response times.

While prior work has explored precomputing KV caches for static input documents~\cite{lu2024turborag}, in contrast, precomputing likely user queries together with their full responses constitutes a conceptually distinct strategy. 
Rather than storing intermediate computational states, this approach performs the entire inference process in advance. 
As a result, it exhibits fundamentally different efficiency characteristics and deployment possibilities. 
In particular, it enables both significantly lower latency for user queries and static deployment on edge devices without GPU requirements.

Beyond performance gains, this precomputed query approach offers architectural flexibility by decoupling heavy inference from user-facing systems.
For instance, responses can be precomputed offline using powerful LLMs, while lightweight models on edge devices handle user interactions.
This hybrid architecture delivers high-quality outputs without requiring real-time inference by large models, making it well-suited for resource-constrained environments.

\subsection{Opportunities for Leveraging Storage} 
GPU-centric computational architecture in LLM inference enables rapid responses but simultaneously creates issues of high cost and memory limitations.
Despite various optimization techniques being proposed, as computational burden increases, GPU resources become more saturated, and existing memory-based offloading struggles to efficiently store and reuse large-scale data.
In this context, storage-based approaches present important motivating opportunities that can address these challenges.
First, \textbf{Reducing compute costs:}
Loading precomputed results from storage reduces redundant GPU computation for similar queries.
Second, \textbf{Enhancing scalability:}
While memory offloading is capacity-constrained, storage provides a scalable and cost-efficient alternative for retaining large volumes of precomputed or cached data.
Third, \textbf{Reducing response latency:}
By precomputing responses to frequently occurring queries or contexts offline and storing them in storage, they can be immediately reused when similar queries are received.

\section{Our Approach}
\label{sec:design}

\subsection{Design Overview}
\begin{figure}[t]
\centering
\includegraphics[scale=0.68]{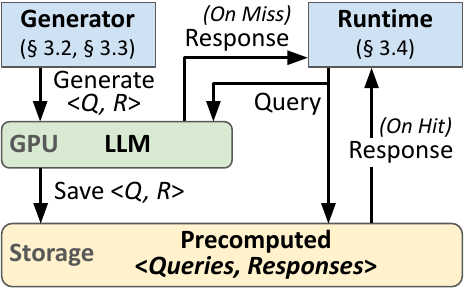}
\caption{\sys{} system architecture.}
\label{fig:fig_overview}
\end{figure}

In this section, we introduce \sys{}, a system designed to precompute query-response pairs for efficient LLM inference.
Figure~\ref{fig:fig_overview} provides an overview of \sys{}’s system, which consists of two main components: {\it Generator} and {\it Runtime}.

{\bf Generator (Offline).}
The Generator leverages LLM to produce user-like queries from an existing knowledge base, prioritizing those likely to occur in real-world scenarios.
This LLM-driven generation ensures the queries are natural and contextually relevant.
It focuses on maximizing coverage while avoiding redundancy, enabling the precomputed set to span a broad range of potential queries.
Each query–response pair is stored in a vector database with disk-based indexing to support fast, similarity-based retrieval.
By shifting query generation and computation into an offline phase, \sys{} efficiently leverages both storage and GPU resources.

{\bf Runtime (Online).}
When a new user query arrives, \sys{} concurrently performs vector search over the precomputed pairs and initiates LLM inference.
If the system finds a highly similar query in storage, it immediately returns the corresponding precomputed response and terminates the ongoing LLM inference—cutting down on unnecessary GPU computation.
In cases where no sufficiently similar match is found, the query proceeds through normal LLM inference, and its newly generated response can optionally be added to the precomputed set for future reuse.

\subsection{Deduplicated Query Generation}
\label{sec:QRgen}

To maximize the effectiveness of precomputed queries, \sys{} avoids generating redundant entries that add little real benefit.
Such duplication not only increases storage overhead but also degrades retrieval performance as the database scales.
Therefore, ensuring a high hit rate while keeping the vector database focused on diverse, meaningful queries is crucial for low-latency real-time inference.
To address these challenges, \sys{} Generator introduces two techniques: {\it Adaptive Query Masking} and {\it Adaptive Sampling}.

\textbf{Adaptive Query Masking.} 
To suppress redundant query generation, this technique injects recently generated queries into the LLM's context window.
Masking candidates are selected from prior outputs, tokenized, and included only if they fully fit within the remaining token budget—calculated by subtracting the token length of the knowledge chunk and prompt scaffolding from the model’s maximum context length.
This token-level control ensures inclusion of only complete prior queries, preserving logical continuity and reducing semantic repetition.

\textbf{Adaptive Sampling.} 
To encourage semantic diversity, this method dynamically adjusts the generation temperature based on similarity with previously generated queries.
If a newly generated query exceeds a similarity threshold ($S_{th\_Gen}$) with any existing query (set to 0.99 in our implementation), it is discarded to prevent redundancy in the embedding space.
Note that if $S_{th\_Gen}=1.0$, only queries that are exactly identical would be filtered, while lower values allow filtering of semantically similar ones.
The generation temperature is initialized at 0.7—a widely used default in prior implementations~\cite{victoria_temperature}—to promote coherent yet knowledge-grounded outputs.
As similar queries appear more frequently, the temperature is increased by 0.1 each time, up to a maximum of 1.0.
While various sampling parameters can be used to control output diversity, many rely on discrete filtering or heuristics that are often hard to tune and interpret.
In contrast, temperature offers a simple and continuous means of modulating the entropy of the softmax distribution by directly scaling the logits.
This makes it particularly suitable for staged generation scenarios where gradual control over diversity is desirable.

\subsection{Response Generation}
Each query is treated as a potential user input and processed through the same LLM inference pipeline used in production, including prompt templating, system message injection, and output formatting.
To further improve response quality, \sys{} supports the use of higher-capacity or more capable LLMs during the precomputation phase.
This ensures that the stored responses are of higher quality than what might be achievable using smaller or resource-constrained models typically used in real-time, on-device environments.

The precomputed query–response pairs are stored in a vector database. 
Each query is embedded into a high-dimensional vector and indexed to enable efficient similarity search, while the corresponding response is stored as metadata associated with the vector.
By eliminating the need for real-time computation, this design reduces response latency and significantly lowers serving costs.
Moreover, by leveraging a stronger LLM at precompute time, the system achieves both high-quality response generation and the efficiency of lightweight inference.
This design ensures a consistent user experience even in resource-constrained environments.

\begin{figure}[t]
\centering
\includegraphics[scale=0.55]{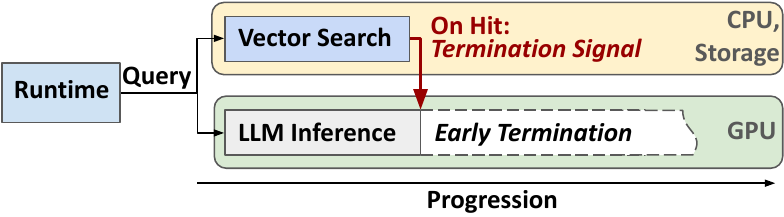}
\caption{\sys{} Runtime executes vector search and LLM inference in parallel, sending a termination signal upon a query hit.}
\label{fig:fig_parallel}
\end{figure}

\subsection{Parallel Execution}
\label{sec:parallel_exec}
As illustrated in Figure~\ref{fig:fig_parallel}, when a user query is received, the \sys{} Runtime concurrently executes vector search and LLM inference.
The vector search efficiently retrieves similar precomputed queries from the vector database.
This concurrency is possible because vector search utilizes CPU and storage resources, whereas LLM inference runs independently on GPU, allowing both operations to proceed without resource contention.
If the similarity exceeds a runtime threshold ($S_{th\_Run}$), the matched precomputed response is immediately returned.
Simultaneously, a termination signal is sent to cancel the ongoing LLM inference, avoiding redundant computation and reducing latency.
If no suitable match is found, the LLM inference continues to generate a new response.
The parameter $S_{th\_Run}$ affects the trade-off between the hit rate of precomputed queries and the overall response quality.
A lower $S_{th\_Run}$ increases the hit rate by allowing matches with less similar queries; 
however, such matches may produce responses that are less aligned with the user query, potentially degrading response quality.
Conversely, a higher $S_{th\_Run}$ reduces the hit rate but can improve response quality by ensuring higher similarity.
We analyze the impact of this trade-off in detail in Section~\ref{sec:eval}, along with Table~\ref{tab:threshold_performance}.

\color{black}

\section{Evaluation}
\label{sec:eval}

{\bf System Environment:} All experiments are conducted on a machine equipped with two Intel(R) Xeon(R) Gold 6442Y processors (96 cores), an NVIDIA H100 GPU (80~GB), 2~TB RAM, and 3.84~TB PCIe Gen5 NVMe storage.

{\bf Datasets:} We evaluate our method on three standard QA datasets: SQuAD~1.1~\cite{rajpurkar2016squad}, NarrativeQA~\cite{kocisky2018narrativeqa}, and TriviaQA~\cite{joshi2017triviaqa}.
Varying context lengths across datasets allow evaluation under different retrieval and reasoning conditions.
We randomly sampled 200 documents from each dataset and generated 150K precomputed query-response pairs.

{\bf Models and Tools:} For query-response generation and baseline LLM inference, we adopt the LLaMA~3.1~8B~\cite{grattafiori2024llama3}, and vector search is implemented with DiskANN~\cite{NEURIPS2019_09853c7f}.
For dense retrieval, we utilize the embedding model ALL-MiniLM-L6-v2~\cite{all-minilm-l6-v2}.
Inference is carried out using vLLM~\cite{kwon2023vllm, vllm_code}.

\textbf{Quality Metrics.}
We evaluate response quality using three widely used metrics:
Unigram~\cite{van1979information}, ROUGE-L~\cite{lin-2004-rouge}, and BERT Score~\cite{zhang2020bertscore}.
Unigram measures lexical overlap via the harmonic mean of precision and recall.
ROUGE-L captures structural similarity through the longest common subsequence.
BERTScore assesses semantic similarity based on contextual embeddings.

\begin{figure}[t]
\centering
\includegraphics[scale=0.35]{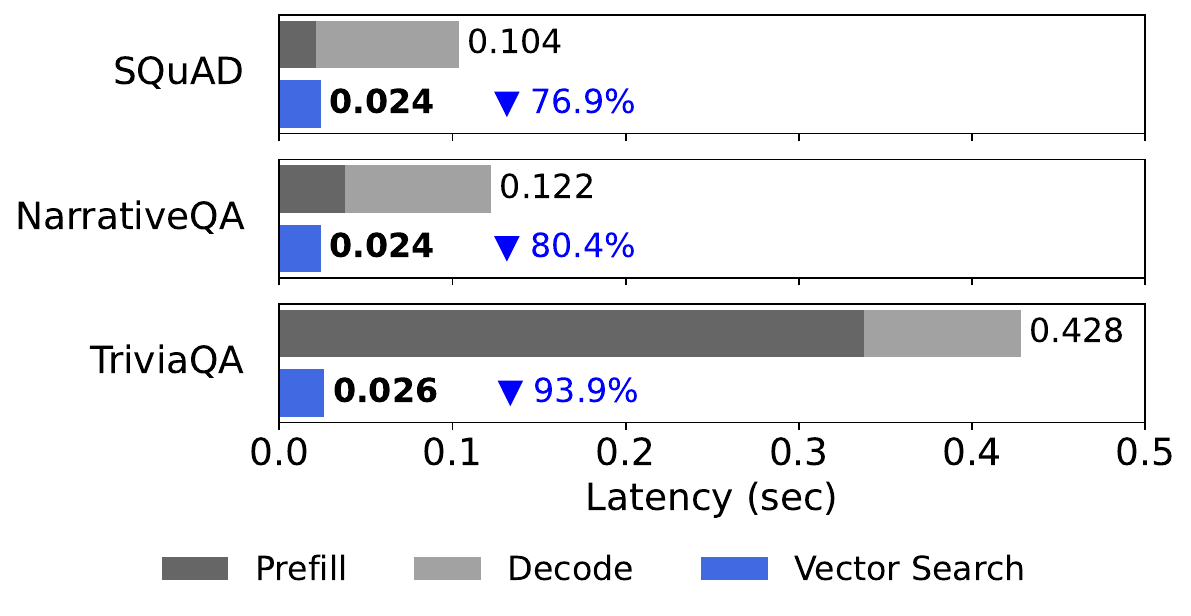}
\caption{Response latency of traditional LLM inference vs. vector search in \sys{} across different datasets.}
\label{fig:latency_comparison}
\end{figure}

When a user query is received, the \sys{} Runtime responds by retrieving the result from the set of precomputed queries.
If a matching precomputed query is found, the response is returned with only vector search latency, bypassing LLM inference.
Figure~\ref{fig:latency_comparison} compares the vector search latency with that of traditional LLM inference across three datasets.
The results indicate that LLM inference latency increases with the growth of context size appended to the knowledge base, whereas vector search time remains stable at approximately 0.02 seconds across different datasets.
Vector search yields an average 8.6× speedup over full LLM inference, with significant latency reductions on precomputed query hits.
Existing prefix cache approaches can only optimize the prefill phase, while the latency of the decode phase remains unaffected.
Even if prefix caching approaches entirely eliminate the latency of the prefill phase, our method remains on average 3.5× faster than the latency incurred by the decode phase alone.

In real-world scenarios, the overall response latency is heavily influenced by how frequently user queries hit the set of precomputed queries.
In practice, the overall response latency depends largely on the frequency with which user queries match entries in the precomputed set.
We refer to this as the effective latency, which we define as follows:
\hfill \(\textit{effective\_latency} = \textit{hit\_rate} \times \textit{vector\_search\_latency} + \textit{miss\_rate} \times \textit{llm\_inference\_latency}\).
\sys{} is designed to eliminate additional runtime overhead on query misses by leveraging parallel execution, ensuring that latency in such cases remains equivalent to traditional LLM inference.
In Table~\ref{tab:storinfer_grouped_latency}, we configure the $S_{th\_Run}=0.9$ such that our approach achieves response quality comparable to that of traditional LLM inference.
Notably, the table also shows that using precomputed query storage improves response latency without compromising quality, validating \textbf{RQ1} – latency can be reduced while preserving response quality.

\begin{table}[t]
\small
\centering
\caption{Hit rate and effective latency by dataset and query generation approach. (\%) indicates latency reduction relative to traditional LLM inference.}
\vspace{0.5cm}
\begin{tabular}{lrr}
\hline
\textbf{Dataset} & \textbf{Hit Rate} & \textbf{Effective Latency} \\
\hline
\multicolumn{3}{l}{\textbf{SQuAD}} \\
\quad Random & 0.180  & 0.090~(\ensuremath{\blacktriangledown}13.8~\%) \\
\quad Deduplicated  & 0.225 & 0.086~(\ensuremath{\blacktriangledown}17.3~\%) \\
\hline
\multicolumn{3}{l}{\textbf{NarrativeQA}} \\
\quad Random & 0.080  & 0.114~(\ensuremath{\blacktriangledown}6.4~\%) \\
\quad Deduplicated  & 0.110 & 0.111~(\ensuremath{\blacktriangledown}8.8~\%) \\
\hline
\multicolumn{3}{l}{\textbf{TriviaQA}} \\
\quad Random & 0.050  & 0.408~(\ensuremath{\blacktriangledown}4.7~\%) \\
\quad Deduplicated  & 0.080 & 0.396~(\ensuremath{\blacktriangledown}7.5~\%) \\
\hline
\end{tabular}
\label{tab:storinfer_grouped_latency}
\end{table}

\begin{figure}[t]
\centering
\includegraphics[scale=0.3]{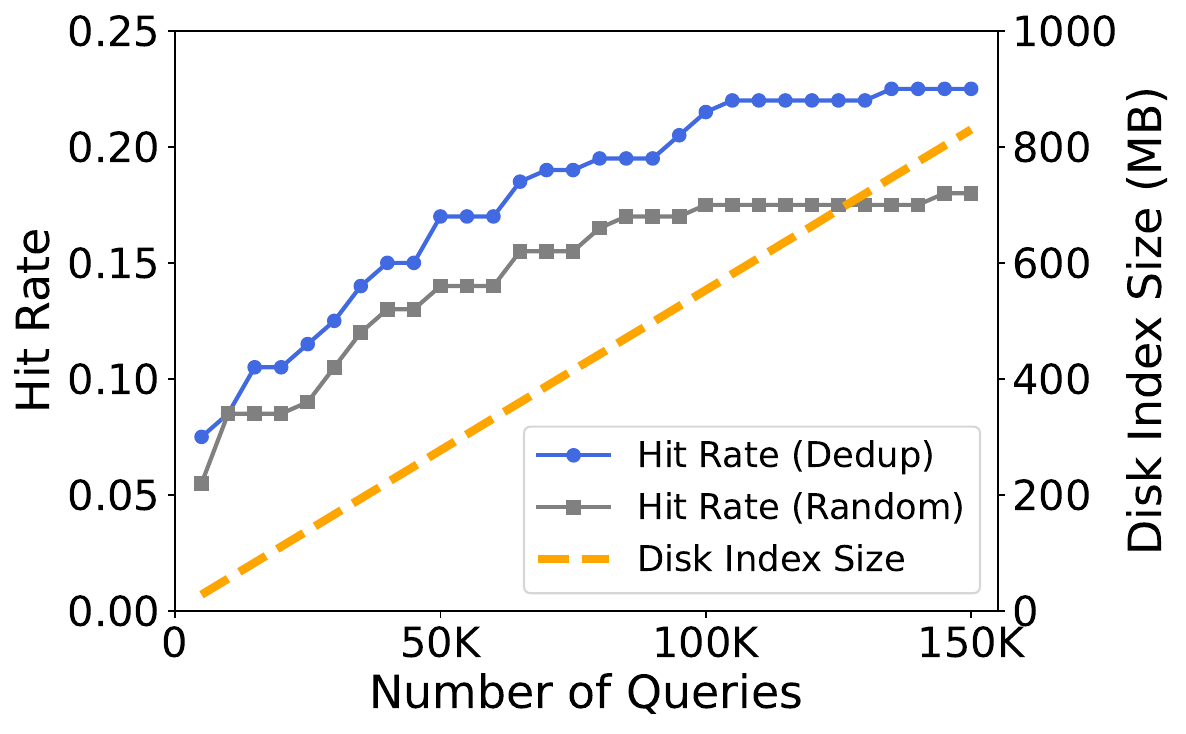}
\caption{Hit rate and storage usage with increasing number of precomputed queries on the SQuAD dataset.}
\label{fig:hit_rate}
\end{figure}

To evaluate the effectiveness of our deduplicated query generation, we compare its hit rate and effective latency against that of a baseline approach in which queries are generated randomly without deduplication.
In Table~\ref{tab:storinfer_grouped_latency}, the deduplicated approach yields a higher hit rate and lower effective latency compared to the random baseline.
Figure~\ref{fig:hit_rate} demonstrates that as the number of precomputed queries increases, our deduplicated query generation yields progressively higher coverage (hit rate) compared to the random approach, thereby validating \textbf{RQ2} – we can generate diverse queries to anticipate user needs.
Figure~\ref{fig:hit_rate} also shows that increasing the number of stored queries (and consequently storage usage) leads to a higher hit rate, thus validating \textbf{RQ3} – highlighting that increased storage usage can lead to improved effective latency.

To ensure the reliability of our response quality evaluation, we employ three different metrics.
Table~\ref{tab:threshold_performance} summarizes the effect of varying runtime thresholds on response quality and hit rate.
All precomputed response pairs in \sys{} are generated using LLaMA~3.1~8B, and we use both LLaMA~3.1~8B and LLaMA~3.2~1B—with different parameter scales—as baselines for comparison.
When the similarity threshold ($S_{th\_Run}$) is set to 0.9, the response quality closely matches LLaMA~3.1~8B.
As $S_{th\_Run}$ decreases, responses are matched to less similar precomputed queries, which leads to a drop in overall response quality.
However, this comes with a dramatic increase in hit rate.
Notably, at $S_{th\_Run} = 0.5$, the response quality surpasses that of LLaMA~3.2~1B, while achieving a hit rate of 93\%.
This suggests that in resource-constrained environments—such as on-device applications using small LLMs—\sys{} can deliver guaranteed response quality with significantly reduced response latency by leveraging high-quality precomputed responses.

\color{black}

In our experimental setup, generating a single precomputed query-response pair takes approximately 0.3 seconds.
Due to deduplication threshold, some queries are discarded, resulting in a maximum generation time of up to 0.6 seconds per pair.
For 150K precomputed queries, the high-dimensional vector index occupies 810~MB of storage, while the associated response metadata requires an additional 20~MB.
In total, using 830~MB of storage, our system maintains response quality, achieves up to a 22.5\% hit rate, and reduces latency by 17.3\%.

\begin{table}[t]
    \small
    \centering
    \caption{Effect of runtime threshold on response quality and hit rate for precomputed queries on the SQuAD.}
    \vspace{0.5cm}
    \begin{tabular}{c|ccc|cc}
        \toprule
        \multirow{2}{*}{\raisebox{-.7\totalheight}{\shortstack{Quality\\Metric}}} & \multicolumn{3}{c|}{$S_{th\_Run}$} & \multicolumn{2}{c}{LLaMA} \\
        \cmidrule(lr){2-4}\cmidrule(lr){5-6}
        & 0.5 & 0.7 & \textbf{0.9} & \textbf{8B} & 1B \\
        \midrule
        Unigram F1 & 0.389 & 0.446 & {\bf 0.570} & {\bf 0.589} & 0.307 \\
        ROUGE-L F1 & 0.404 & 0.463 & {\bf 0.586} & {\bf 0.598} & 0.332 \\
        BERT F1 & 0.308 & 0.353 & {\bf 0.458} & {\bf 0.439} & 0.305 \\
        \midrule
        {\it Hit Rate} & {\it 0.930} & {\it 0.690} & {\it 0.225} & -- & -- \\
        \bottomrule
    \end{tabular}
    \label{tab:threshold_performance}
\end{table}
\section{Conclusion}
\label{sec:conclusion}

We presented \sys{}, a storage-assisted approach that accelerates LLM inference by leveraging precomputed query –response pairs. To the best of our knowledge, this is the first approach that integrates precomputed query–response pairs into LLM inference workflows. Our evaluation across multiple QA benchmarks demonstrates that \sys{} achieves up to 17.3\% latency reduction.

\section*{Acknowledgments}
The authors would like to thank SMRC (Samsung Memory Research Center) for providing the infrastructure for this work~\cite{smrc}.

\bibliographystyle{plain}
\bibliography{references}

\end{document}